# Schlieren texture induced Anderson localization in an organic exciton-polariton laser

*Florian Le Roux[1*], Andreas Mischok[1], Francisco Tenopala-Carmona[1], Malte C. Gather[1,2*]*

[1] Humboldt Centre for Nano- and Biophotonics, Department of Chemistry, University of Cologne, Greinstr. 4-6, 50939 Köln, Germany

[2] Organic Semiconductor Centre, SUPA School of Physics and Astronomy, University of St Andrews, St Andrews, KY16 9SS UK

*correspondence to: malte.gather@uni-koeln.de, florian.leroux@uni-koeln.de

## Abstract:

Non-linearities in organic exciton-polariton microcavities represent an attractive platform for second-generation quantum devices. However, progress in this area hinges on the development of material platforms for high-performance polariton lasing, scalable and sustainable fabrication, and ultimately strategies for electrical pumping. Here, we show how introducing Schlieren textures in a liquid crystalline conjugated polymer and the associated microdomains of distinct chain orientation enable in-plane Anderson localization of polaritons. In high-*Q* distributed Bragg reflector microcavities, this strong localization facilitated polariton lasing at unprecedented thresholds of 136 fJ per pulse, thus providing a pathway to the study of fundamental effects at low polariton numbers. Anderson localization further permitted polariton lasing in more lossy metallic microcavities while maintaining a competitive lasing threshold. The facile fabrication of these cavities will drastically reduce the complexity of integrating polariton laser with other structures and the high conductivity of metallic mirrors provides a route to electrical pumping.

**Introduction**

Frenkel excitons in organic semiconductors have large binding energies and oscillator strengths[1,2]. When hybridized with the photonic mode of a planar microcavity, they form exciton-polaritons that are stable at room-temperature and that show a large Rabi splitting, $\hbar\Omega$, between the resulting lower polariton (LP) and upper polariton (UP) branches; values ≥ 1 eV have been demonstrated in organic microcavities[3–6]. Nonlinearities are manifest in these systems during polariton lasing, that is when polaritons scatter to macroscopically occupy the ground state and then decay through emission of coherent photons. Polariton lasing has been observed in a wide range of organic microcavities containing single-crystals of anthracene[7], small organic molecules[8], polymers[9] or proteins[10]; where LP lifetimes have been sufficiently long for efficient thermalization of the LP ground state population, this also enabled out-of-equilibrium Bose Einstein Condensation (BEC).

Reducing the threshold of polariton lasing and BEC is of great importance for fundamental studies and for fascinating applications of these phenomena, e.g. superfluidity of light[11], optical logic at room-temperature[12], and single-photon detection[13], because threshold is often directly linked to device performance and power consumption. High photoluminescence quantum yield (PLQY)[14], fast exciton decay rates[15], and more recently alignment of the exciton transition dipole moment[16,17] have proven advantageous in this context. By using an active layer of a liquid-crystalline conjugated polymer (LCCP) that is macroscopically aligned into its nematic phase, thresholds down to 2.23 pJ of incident pulse energy have been demonstrated[17], thus putting organic polariton lasers on par with the best organic photon lasers based on vertical-cavity surface-emitting designs[18,19].

In recent decades, the polymorphic nature of liquid crystals has garnered significant interest[20–22], leading to extensive endeavors to classify the numerous phases and textures and their corresponding ordering. Nematic liquid crystals have the ability to align their director, that is the average local direction of their long molecular axis, parallel to a substrate, and when this alignment is not homogeneous, a Schlieren texture[20,23] with locally well-defined director orientation can emerge (μm to sub-μm domains). This texture has been recorded in LCCPs[24,25] and can be regarded as an optical medium in which a high degree of disorder arises from an ensemble of highly ordered domains similar in size to the wavelength of visible light, therefore making it a favorable environment for scattering and possibly confining light.

Anderson localization is a hallmark of strongly disordered optical media and occurs when photons become trapped upon random scattering at local refractive index discontinuities[26]. It was found to occur in photonic-crystal waveguides, where deliberate introduction of disorder induced localized cavity modes[27]. It was further harnessed for spectral selection in random nanolasers by local excitation of isolated modes[28]. Recently, Anderson localization has also been achieved for microcavity polaritons in a 2D photonic crystal, using a top-down approach to control disorder by adjusting the in-plane mesa positions of the photonic crystal[29].

Here, we introduce Schlieren textures in films of the LCCP poly(9,9-dioctylfluorene) (PFO) to induce Anderson localization of polaritons via a scalable bottom-up approach and thus further reduce optical losses in polariton lasers. We find that for films of 15% $\beta$-phase PFO, the discontinuities in refractive index at domain boundaries give rise to Anderson localization of polaritons and that the resulting in-plane confinement leads to a drastic improvement in polariton laser performance. This increase in performance allows us to replace the electrically insulating distributed Bragg reflectors (DBRs) that are conventionally used to form the microcavity by highly conductive metallic mirrors. Metallic mirrors were previously unsuitable for polariton lasers due to their lower reflectivity (e.g., ~ 95% reflectivity for silver compared with > 99.9% for DBRs). The deposition of metallic mirrors via thermal evaporation is substantially simpler, gentler, and more rapid (tens of minutes) than the fabrication of DBRs. In addition, metallic mirrors also offer a direct pathway for charge injection in future electrically pumped polariton lasers.

We first fabricate a high-$Q$ polariton laser made of DBRs sandwiching a film of Schlieren textured PFO. Comparing the spatial pattern of emission under polarized excitation to corresponding polarized transmission micrographs reveals that the emission originates from micro-domains with exciton transition dipole moments closely aligned with the pump polarization. For a given polarization, we observe small high-intensity spots, characteristic of strong localization, which undergo a super-linear increase in intensity as the excitation energy increases. Selective excitation of these emission centers facilitates polariton lasing with a threshold down to $P_{\text{th,DBR/DBR}} = 136$ fJ per pulse. This threshold represents a record for both organic polariton lasers and organic vertical surface-emitting photon lasers; it is more than 16-times lower than the best value demonstrated for a cavity containing a macroscopically aligned layer of 15% $\beta$-phase PFO[17] and almost two orders of magnitude lower than for a corresponding isotropic polariton laser. We observe an average sub-threshold LP linewidth of ~ 500 μeV, indicating a long polariton lifetime of $\tau_{LP} > 1$ ps and a one order of magnitude increase in $Q$-factor over non-textured cavities[17] and comparable zero-dimensional cavities fabricated using Gaussian-shaped defects[30]. We utilize the improvement in threshold afforded by the Schlieren texture induced Anderson localization to realize anisotropic polariton lasing in a hybrid metal/DBR cavity with a threshold of $P_{th,Ag/DBR}$ of 2.67 pJ per pulse, as well as in a metal/metal cavity with $P_{th,Ag/Ag}$ = 15.19 pJ per pulse. Despite the metal-induced losses, the performance of the purely metallic polariton laser is on par with current state-of-the-art DBR-based polariton lasers. Careful analysis of the hallmarks of polariton lasing—in particular linewidth and blue-shift—for the different types of mirrors reveals a subtle interplay between lasing threshold, localization and surface roughness.

**Results**

**Structure of polariton lasers**

**Figure 1**a illustrates our protocol for forming Schlieren-textured active layers of nematic phase PFO through heating (160°C for 10 minutes) and rapid quenching to room temperature, and finally solvent (toluene) vapour annealing for 24 hours to systematically induce 15% of the PFO $\beta$-phase. Also shown is a typical polarized optical micrograph of the resulting film. Figure 1b shows the characteristic localized emission inside a microcavity sample and Figure 1c illustrates the proposed emission mechanism in which Anderson localization couples with the vertical feedback provided by the cavity mirrors.

We fabricated and compared three types of cavities, namely DBR/DBR, Ag/DBR and Ag/Ag (Figure 1d, for details on fabrication see *Methods*). Since coherent emission is expected to arise from aligned micro-domains in the cavities and with its polarization along the direction of alignment, the optical thickness of each cavity design was optimized for these conditions using transfer matrix calculations (TMCs, Supporting Information Figure S8). The LP energy at normal incidence ($\theta=0°$) was tuned to match the (0-1) vibronic peak of the emission of $\beta$-phase PFO, which is at 2.67 eV (Supporting Information Figure S1); this strategy has proven beneficial for polariton laser perfomance[15,17].

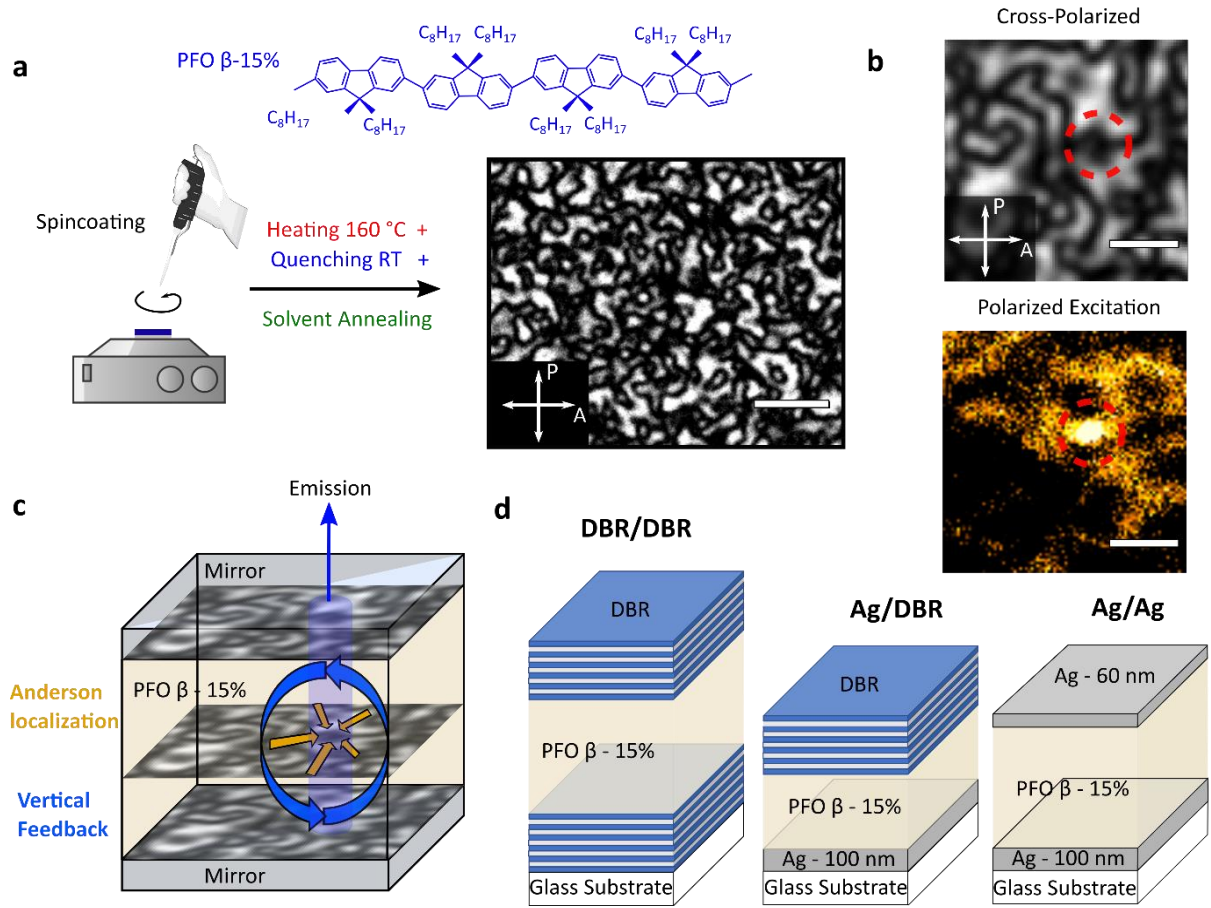

**Figure 1. Fabrication and working principle of the Schlieren-textured polariton lasers.**
**a**, Illustration of protocol for forming Schlieren textured active layers of PFO with 15% β-phase content. Chemical structure of the β-phase chain conformation of PFO shown in blue. Polarized transmission optical micrograph of the resulting film placed between a crossed polarizer (P) and analyzer (A) pair. No light is transmitted when the exciton transition dipole moment lies perpendicular to either the polarizer or analyzer, while maximum transmission occurs when the transition dipole moment lies at 45° relative to both the analyzer and polarizer. **b**, Polarized optical micrograph (top) and corresponding emission under non-resonant polarized excitation (false colour, bottom) on the DBR/DBR cavity surface. The dashed red-circle indicates an area of localized emission. **c**, Proposed working principle of the Schlieren textured polariton laser: Anderson localization arising from local refractive index discontinuities provides in-plane confinement while the cavity mirrors provide vertical feedback. **d**, Schematic structure of the dielectric DBR/DBR, hybrid Ag/DBR and metallic Ag/Ag cavities used in this study. Scale bar in a, 10 μm; in b, 4 μm.

**Strong localization of polaritons under polarized excitation**

**Figure 2** shows the spectrally and spatially resolved emission from a thickness-optimized DBR/DBR cavity under non-resonant pulsed excitation with an extended pump spot (diameter $d_{pump}$ ~ 150 µm; wavelength 355 nm / photon energy 3.49 eV; pulse duration 25 ps; repetition rate 250 Hz). The polarization of the pump spot was set to either vertical ($\phi_{pump} = 0°$; Figure 2a, b, c) or horizontal ($\phi_{pump} = 90°$; Figure 2d, e, f) and the pulse energy was varied between 500 pJ and 2.5 nJ. Irrespective of pulse energy, we found the spatially resolved emission to be complementary for the two orthogonal polarizations, i.e. areas that appeared bright for $\phi_{pump} = 0°$ tended to be dark for $\phi_{pump} = 90°$ (Supporting Information Figure S4). We attribute this to the fact that parallel or close to parallel alignment of the local transition dipole moment $\boldsymbol{\mu}$ with pump polarization $\boldsymbol{E}$ leads to a larger dot product $\boldsymbol{\mu} \cdot \boldsymbol{E}$ and thus stronger absorption, resulting in brighter emission.

At a pump pulse energy of 500 pJ (2.84 µJ.cm$^{-2}$, Figure 2a, d), the spectra for orthogonal pump polarizations were similar and resembled the LP emission spectra previously observed for an aligned PFO 15% β-phase cavity operating in the sub-threshold regime[17]. As a measure of polariton localization, we calculated the inverse participation ratio (IPR)[29] from the spatially resolved emission, using $\text{IPR} = N\left(\sum_{n=0}^{N} I_n^2\right)/\left(\sum_{n=0}^{N} I_n\right)^2$ where $I_n = \int_{\text{spot } n} |\psi|^2 d\boldsymbol{r}_n$ is the intensity of the *n*-th emission spot and $\psi$ is the emitted field (IPR = 1 in the absence of localization and increases as the emission becomes localized). The average IPR at an excitation of 500 pJ for $\phi_{pump} = 0°$ calculated over 12 disorder realizations (with each disorder realization corresponding to a different region of the sample) was 1.06 (for a given polarization, only pixels with non-negligible emission were counted), a value which demonstrates[29] strong localization. This finding is consistent with 3D-finite-difference-time-domain (FDTD) simulations that also show strong localization of polaritons of the in-plane component of $|\boldsymbol{E}|^2$ (Supporting Information Figure S4 and Figure S5) with a localization length estimated to ~3 µm.

Increasing the excitation pulse energy to 1.5 nJ (8.52 µJ.cm$^{-2}$) led to the formation of high-intensity, strongly localized emission spots, accompanied by high-intensity narrow emission peaks on the red edge of the spectrum (Figure 2b, e). These peaks were confined to the region of LP emission; this behaviour differs from random lasing where emission is typically observed across the entire fluorophore emission spectrum. At an excitation pulse energy of 2.5 nJ (14.2 µJ.cm$^{-2}$), numerous bright spots appeared, and the corresponding spectra were dominated by individual narrow peaks (Figure 2c, f).

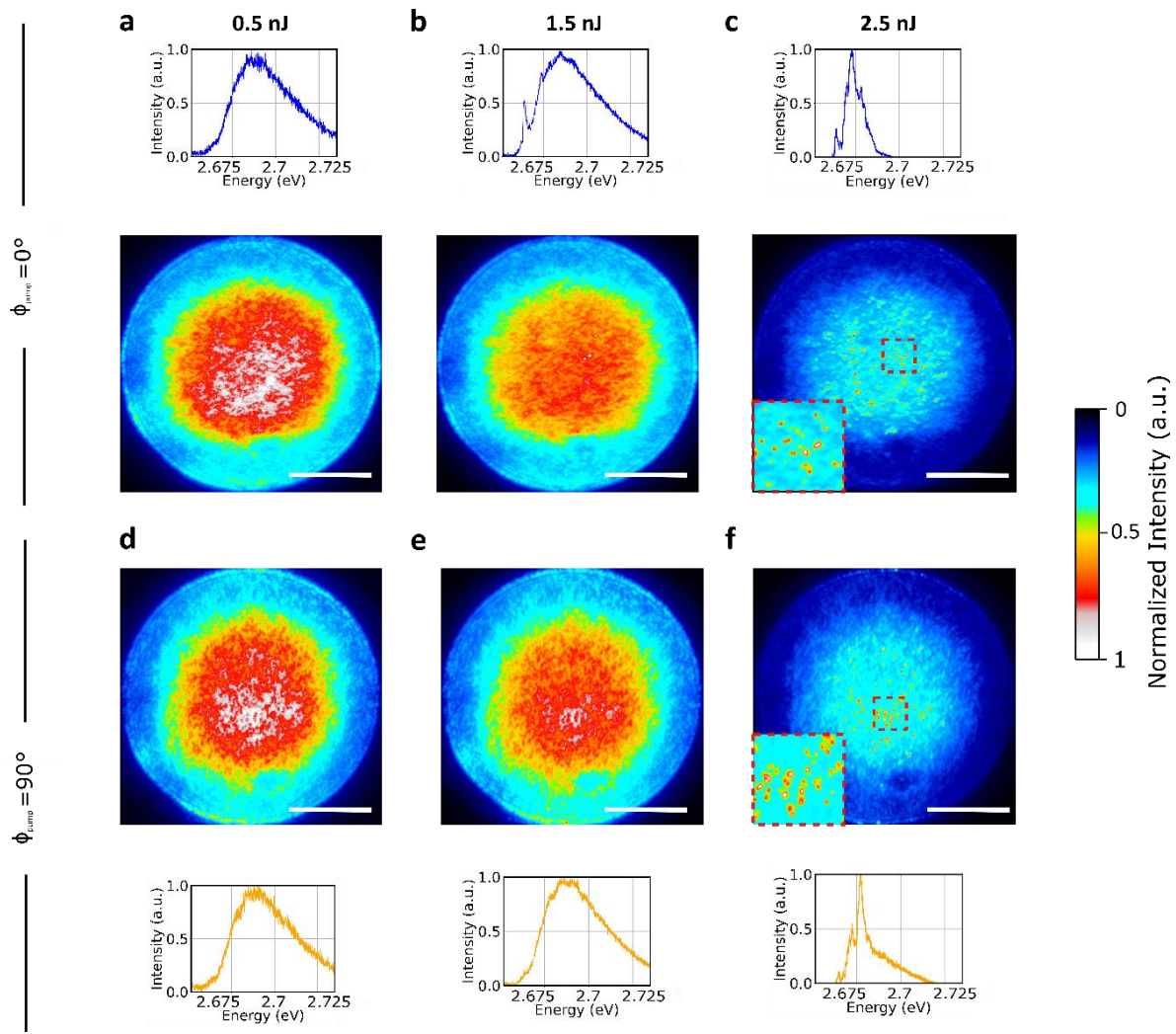

**Figure 2. Spatially and spectrally resolved emission from the Schlieren textured DBR/DBR cavity.** Spatially and spectrally resolved emission from a DBR/DBR cavity with Schlieren textured active layer upon non-resonant excitation with increasing pulse energy and with orthogonal polarizations, $\phi_{pump}= 0°$ (a, b, c) and $\phi_{pump}= 90°$ (d, e, f). Excitation pulse energies were 0.5 nJ (a and d), 1.5 nJ (b and e), and 2.5 nJ (c and f). Note that non-uniform emission is already present at 0.5 nJ, and how high intensity emission spots (in white), accompanied by clear narrow emission peaks in the spectrum, occur from excitation energies of 1.5 nJ. The insets in c and f show magnifications of the areas marked by red-dashed rectangles in the corresponding main panel. (See Supporting Information Figure S2 and Figure S3 for data spanning a wider range of excitation energies.) Scale bars, 50 µm.

**Mode selection and angle-resolved photoluminescence**

Next, we show how focusing the excitation beam to match the localization area ($d_{pump} \sim 3$ μm) enables spectral and spatial selection of strongly localized polariton lasing modes. For this, the angular distribution of the emission from the DBR/DBR cavity was measured via Fourier plane imaging while pumping at $\phi_{pump} = 0°$. When exciting below the polariton lasing threshold, the emission was weak and its angular dispersion was flat (**Figure 3a**), indicating strong confinement of polaritons and resulting mode discretization[31,32]. Above threshold, however, the emission increased nonlinearly, reduced in linewidth to below the resolution of our optical setup (300 μeV), and blue-shifted (Figure 3b). Above threshold, the emission also shows spatial coherence as demonstrated by the appearance of interference fringes on a Michelson interferometer in retro-reflector configuration (Supporting Information Figure S7). These features are hallmarks of the onset of polariton lasing[8,9,33].

Figure 3c shows the angle-resolved emission of a hybrid Ag/DBR cavity at an excitation energy just below the threshold between the linear and super-linear regimes; both a very weak parabolic background emission and an intense flat emission at around 2.665 eV are visible. Above threshold, the flat emission overshadows the background emission (Figure 3d). The spectra of the purely metallic Ag/Ag cavity are similar to the hybrid Ag/DBR cavity, but the excitation pulse energy required to reach threshold is higher and the emission linewidths are broader due to additional losses induced by the second Ag mirror (Figure 3e, f). The weak background emission observed for the Ag/DBR and Ag/Ag cavities is associated with non-localized emission emanating from the surrounding of the localized spot; it is more visible for the Ag/DBR and Ag/Ag cavities as the excitation pulse energies used for these are more than one order of magnitude higher than for the DBR/DBR cavity. For all three cavity types, no emission was observed when the excitation polarization was set to $\phi_{pump} = 90°$ while keeping the same excitation spot; this illustrates how the strong anisotropy of the system enables polarization selection.

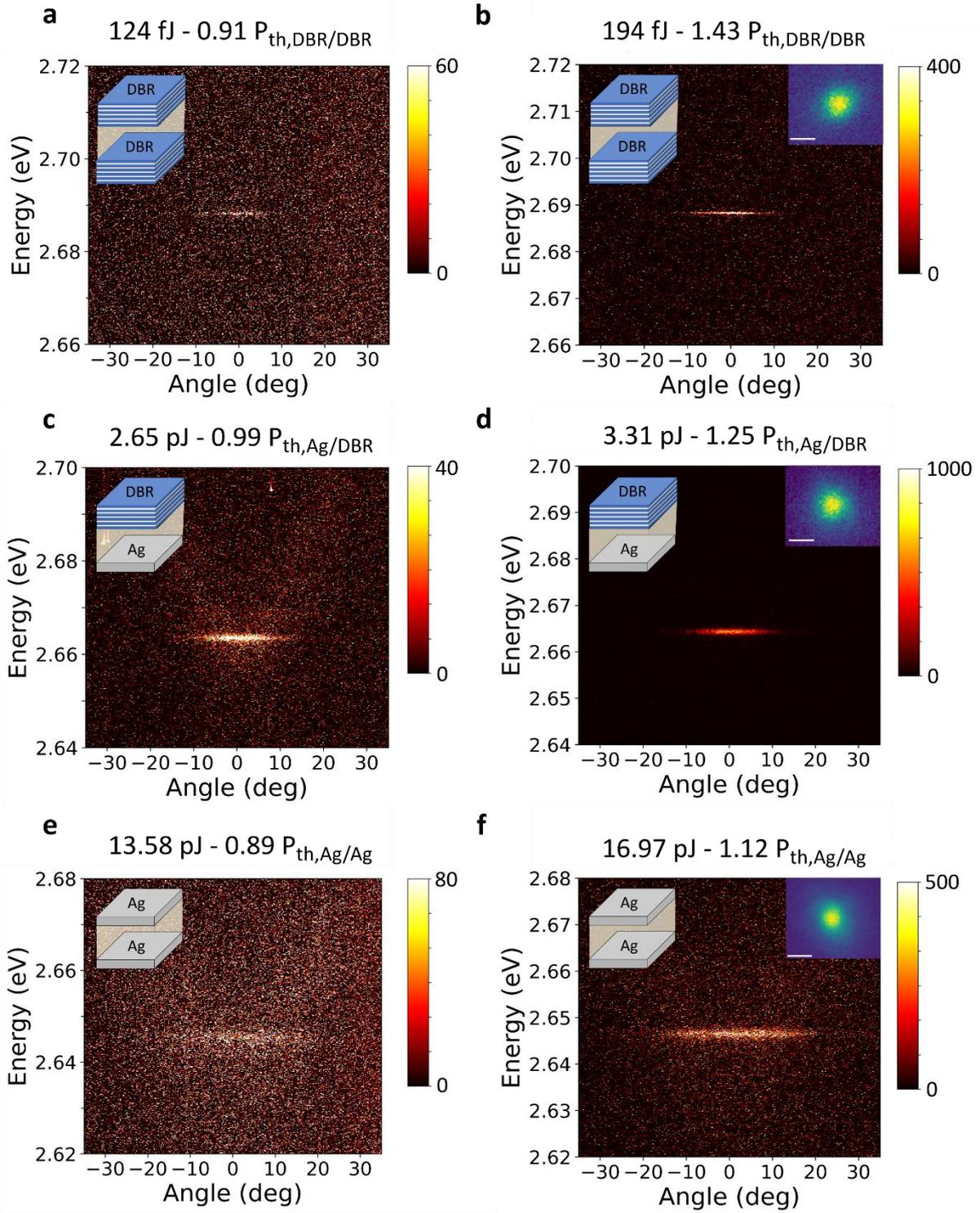

**Figure 3. False colour maps of angle-resolved emission spectra below and above threshold for different device types. a**, **b**, DBR/DBR microcavity. **c**, **d**, Ag/DBR microcavity. **e**, **f**, Ag/Ag microcavity. Excitation just below (a, c, e) and above (b, d, f) the lasing threshold. The insets in (b, d, f) show the corresponding real space emission; scale bars, 2.5 μm. The excitation polarization was vertical ($\phi_{pump} = 0°$). An intensity colour scale is given on the right-hand side of each panel.

**Polariton lasing landmarks**

**Figure 4**a shows the emission intensity versus the incident excitation pulse energy for all samples, in each case integrated over $\theta \in [-2°; 2°]$. The PL spectra used for the integration are shown in Figure 4b-d. For all samples, a clear, super-linear increase in intensity is observed above a specific threshold excitation pulse energy $P_{th}$. The observation of organic polariton lasing from vertical cavities with purely metallic mirrors is particularly noteworthy; due to plasmonic losses, polariton lasers have so far generally relied on DBR mirrors or hybrid metal cavities partially supported by DBRs[34–37].

All three cavity types show a reduction in linewidth and a blueshift of the emission peak, both hallmarks of polariton lasing (Figure 4e-g). The ~ 40 meV reduction in linewidth seen for the Ag/Ag cavity is remarkable; reductions reported in other studies are on the order of a few meV[14,15,30]. In addition, the blueshift of the emission peak increases from ~ 2 meV for the DBR/DBR cavity to ~ 12 meV for the Ag/Ag cavity, which again is more than twice as large as the values reported for other DBR/DBR cavities[14,15,30]. The characteristic blueshifts in organic polariton condensates were recently shown[33] to originate from quenching of the Rabi-splitting:

$$\hbar\Omega_R = \hbar\Omega_0 \sqrt{1 - \frac{2(n_x + n_p)}{n_0}} \quad (1)$$

where $\hbar\Omega_0$ is the vacuum Rabi-splitting, $n_x$ the number of excitons, $n_p$ the number of polaritons and $n_0$ the total number of molecules participating in strong coupling, as the Pauli-blocking principle prevents occupied states from being filled twice (a counter-intuitive observation that stems from the non-bosonic nature of excitons at low excitation densities). The increase in the peak blueshift observed from DBR/DBR ≤ Ag/DBR ≤ Ag/Ag is thus a result of both the increase in $n_x + n_p$ due to the increase in polariton lasing threshold and the increase in $\hbar\Omega_0$ due to the use of metallic mirrors[6]. Containing the electric field more fully inside the active layer with such mirrors is also a desirable lasing feature that has only been observed for a bound-state in the continuum (BIC) so far[38].

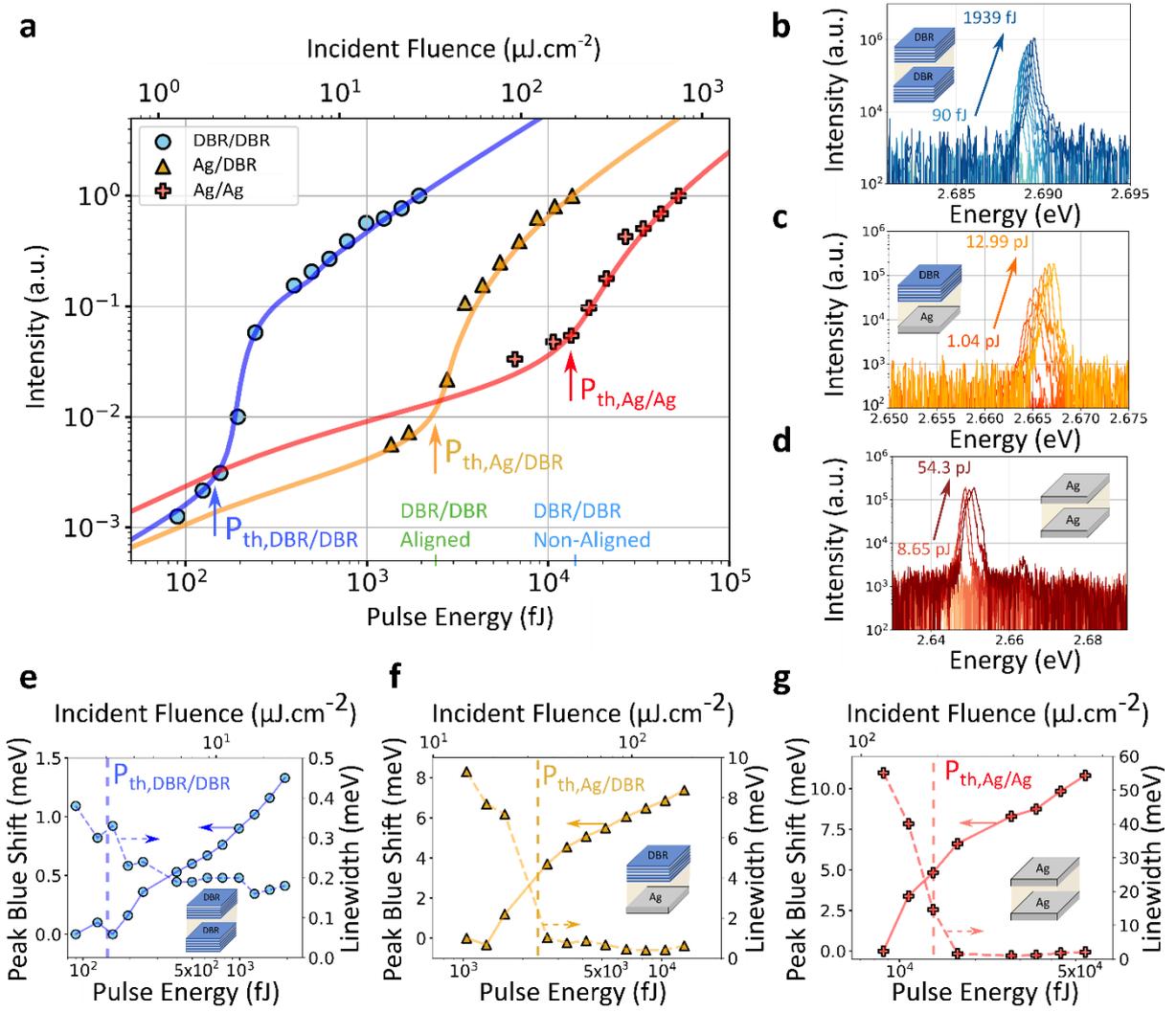

**Figure 4. Analysis of nonlinear intensity increase, blueshift and linewidth reduction of the emission for the different cavity types. a**, Integrated emission intensity versus incident excitation pulse energy and incident fluence for the DBR/DBR (blue symbols), Ag/DBR (orange symbols), and Ag/Ag (red symbols) cavities. Arrows indicate the corresponding polariton lasing thresholds as determined by fitting the measured emission intensity to the kinetic model. The short vertical lines on the x-axis indicate the polariton lasing threshold pulse energies for an aligned DBR/DBR (green) and an isotropic DBR/DBR (blue) cavity without Schlieren texture (values taken from Ref. 17). **b**, **c**, **d**, Emission spectra collected over an angular range $\theta \in [-2°; 2°]$ for increasing excitation pulse energy, transitioning from spontaneous LP emission to polariton lasing, for the DBR/DBR (b), the Ag/DBR (c), and the Ag/Ag (d) cavity. **e**, **f**, **g**, LP linewidth (dashed line) and blue-shift of LP peak (solid line) for the DBR/DBR (e), the Ag/DBR (f), and the Ag/Ag (g) microcavity. The dashed vertical lines indicate the polariton lasing thresholds for each cavity as determined in a. In all cases the excitation polarization was set to $\phi_{\text{pump}} = 0°$.

**Lasing performance and coherence**

The dependence of integrated emission on excitation energy was analyzed further by fitting a kinetic model for the time evolution of the populations of reservoir excitons, $n_R$, and LP, $n_{LP}$, to our data. The master equations for our system are

$$\frac{dn_R}{dt} = \left(1 - \frac{n_R}{N_0}\right) P(t) - \frac{n_R}{\tau_R} - k_B n_R^2 - \frac{W_{ep}}{d} n_R n_{LP} \qquad (2)$$

$$\frac{dn_{LP}}{dt} = W_{ep} n_R n_{LP} - \frac{n_{LP}}{\tau_{LP}} + d \cdot f \frac{n_R}{\tau_R} \qquad (3)$$

where $N_0$ is the excitation density in the PFO film[39] (~7 x 10$^{20}$ cm$^{-3}$), $\tau_R$ is the reservoir exciton lifetime (measured at 450 ps, in-line with previous reports[40]), $d$ is the active layer thickness, and $\tau_{LP}$ is the LP lifetime. $P(t)$ is the temporal profile of the pump pulse, approximated as $P(t) = 2P_{eff} \exp(-t^2(2\sigma^2)^{-1})$ with $FWHM = 2\sqrt{2ln2}\sigma = 25$ ps and $P_{eff} = P_0(dS\hbar\omega_{pump})^{-1}$ where $S$ is the size of the excitation spot, $\hbar\omega_{pump} = 3.49$ eV, $P_0$ is the pump pulse energy, and the factor 2 in front of $P_{eff}$ accounts for enhanced absorption due to alignment of the exciton transition dipole moments and pump polarization[16,17]. The free parameters for non-textured cavities, e.g. a macroscopically aligned layer of 15% $\beta$-phase PFO, are the exciton bimolecular annihilation rate, $k_B$, the exciton reservoir-to-LP resonant scattering rate, $W_{ep}$, and the fraction of spontaneous scattering from the exciton reservoir to the LP, $f$. The latter can be regarded as the equivalent of the $\beta$-factor in conventional photonic lasers, yet its interpretation in lossy systems remains challenging[15]. An estimate for $\tau_{LP}$ is obtained from the linewidth of sub-threshold LP emission at $\theta = 0°$ and is referred to as $\tau_{LP-exp}$.

The values of $k_B$ and $W_{ep}$ fitted for a cavity with a macroscopically aligned layer of 15% $\beta$-phase PFO (Supporting Information Section III, Figure S6) are then taken as fixed parameters for the Schlieren cavities in which $\tau_{LP-fit}$ and $f$ become the only free parameters in order to relate the reduction in threshold to the narrow emission linewidths measured in the linear regime. **Table 1** summarizes the laser performances for the DBR/DBR, Ag/DBR and Ag/Ag cavities and the corresponding kinetic parameters.

The radiative decay times fitted with the kinetic model ($\tau_{\text{LP-fit}}$) are in close agreement with the decay times extracted from the linewidth of the LP emission below threshold ($\tau_{\text{LP-exp}}$). The spontaneous scattering fraction, $f$, increases with the introduction of the Ag mirrors; $f_{\text{DBR/DBR}} = 0.03$, $f_{\text{Ag/DBR}} = 0.17$, and $f_{\text{Ag/Ag}} \approx 1$.

**Table 1.** Parameters extracted from the kinetic model and polariton lasing performance for the DBR/DBR, the Ag/DBR, and Ag/Ag cavities.

| Sample | LP radiative decay time from peak width $\tau_{\text{LP-exp}}$ [fs] | Resonant polariton scattering rate $W_{ep}$ [cm$^3$s$^{-1}$] | Spontaneous scattering fraction $f$ | Bimolecular annihilation rate $k_B$ [cm$^3$s$^{-1}$] | LP radiative decay time from kinetic model $\tau_{\text{LP-fit}}$ [fs] | Incident threshold pulse energy $P_{th}$ |
|---|---|---|---|---|---|---|
| DBR/DBR | 1320 | 2.6 x 10$^{-6}$ | 0.03 | 7.5 x 10$^{-9}$ | 660 | 136 fJ |
| Ag/DBR | 83 | 2.6 x 10$^{-6}$ | 0.17 | 7.5 x 10$^{-9}$ | 55 | 2.67 pJ |
| Ag/Ag | 16.7 | 2.6 x 10$^{-6}$ | $\approx 1$ | 7.5 x 10$^{-9}$ | 26 | 15.2 pJ |

By fitting the experimental data to the kinetic model, a polariton lasing threshold of $P_{\text{th,DBR/DBR}} = 136$ fJ is obtained for the DBR/DBR cavity. This is the lowest threshold pulse energy reported for any organic polariton or vertical cavity surface emitting photon laser to date. It is more than 16 times lower than the previous record[17], achieved for a DBR/DBR cavity with an active layer of macroscopically aligned 15% β-phase PFO ($P_{\text{th,Aligned}} = 2.23$ pJ).

Relative to a DBR/DBR cavity with an active layer of macroscopically aligned 15% β-phase PFO, the $Q$-factor of the cavity with the Schlieren textured active layer increased approximately 10-fold ($Q_{\text{Aligned-exp}} \sim 530$ vs $Q_{\text{DBR/DBR-exp}} \sim 5300$; values estimated from the width of the subthreshold LP spectrum of the respective cavities). We attribute this to a combination of two effects: Firstly, Anderson localization leads to a direct increase in $Q$-factor (~2-fold increase)[28] and secondly, the resulting reduction in mode volume further protects the localized mode from losses due to sample roughness (leading to a ~4-5-fold increase in $Q$-factor, see Supporting Information Section VI). By contrast, the Ag/DBR cavity only shows the direct ~2-fold Q-factor enhancement mediated by Anderson localization. This is consistent with our TMCs, which show that upon introduction of the bottom Ag mirror, metallic losses become dominant over losses induced by to the ~1 nm sample roughness (Supporting Information Section VI). Losses increase further when the top Ag mirror is evaporated directly on the rough polymer layer as the metal film formation is prone nanocluster formation.

**Discussion**

We demonstrated how inducing a Schlieren texture in an LCCP cavity via a simple heating and quenching step can drastically enhance polariton lasing performance. Anderson localization of anisotropic polaritons allowed polariton lasing with a threshold down to $P_{\text{th,DBR/DBR}} = 136$ fJ per pulse for our DBR/DBR cavity, which represents a two orders of magnitude improvement compared to a similar device using an isotropic active material, and a 16-fold improvement over the previous record[17]. The threshold pulse energy in our system corresponds to an incident pulse fluence of $F_{\text{inc}} = 1.94$ uJ.cm$^{-2}$. A direct comparison of this value to the threshold fluence of polariton lasers with homogeneous active layers is however not particularly meaningful due to the strongly reduced mode volume in our Schlieren textured device. In general, whether the pulse fluence or the absolute pulse energy at threshold is the more important figure of merit, depends on the intended application of the laser. However, many future applications of polariton lasers will benefit from having the lowest possible threshold pulse energy. For instance, the low threshold pulse energy of our polariton laser implies that at threshold we create as few as ~300k polaritons within the cavity, which represents a crucial advance towards the observation of quantum effects at low polariton numbers[41,42]. In addition, the narrow LP linewidth of ~500 μeV in the DBR/DBR cavity suggests a LP lifetime of >1 ps, which could be used to coherently address polaritons before their decay through the DBR mirrors.

The dramatic improvement in lasing performance enabled by the Schlieren texture further permitted the fabrication of polariton lasers with conductive metallic mirrors. Use of metallic mirrors is often considered a prerequisite for efficient current injection in future electrically driven organic polariton lasers but has long been considered inaccessible due to the high optical losses associated with them. Recently, some notable progress has been made on metal cavity polariton lasers, with demonstrations of thermalization of the LP in a hybrid DBR-protein-metal cavity[34] and lasing in a metal-metal cavity containing thick single-crystal layers (thickness > 2 μm)[43]. However, these structures exhibited either a high threshold (8 nJ per pulse[34]) or involved challenging active layer fabrication[43].

Future studies on fundamental aspects of polariton lasing as well as the development of new applications based on these phenomena will benefit from our novel bottom-up device platform in which changes in photonic environment allow to tune blueshift over one order of magnitude (from 1 to 10 meV) and giant reductions in linewidth are accessible (up to 40 meV). We anticipate that the wide breadth of LCCPs and recent advances offered by patternable alignment layers (using either photomasks[44] or direct two-photon laser writing[25]) will allow to reproduce and optimize the features of the Schlieren texture, e.g., to explore topologic phenomena[45], further optimize performance, and add new functionalities such as polarization sensitivity to polariton based single photon detectors[13].

**Online Methods**

*Materials*: PFO was supplied by the Sumitomo Chemical Company, Japan and used as received. The peak molecular weight was $M_{pPFO} = 50 \times 10^3$ g mol$^{-1}$. For the dielectric mirrors, Ta$_2$O$_5$ and SiO$_2$ were sputtered from >99.99% oxide targets (Angstrom Engineering). For the metal mirrors, silver and aluminium pellets (99.99%, Kurt J. Lesker company) were thermally evaporated. The substrates used were display-grade glass (Eagle XG, Howard Glass), sized 24 mm x 24 mm.

*Microcavity fabrication:* The DBR/DBR cavity contains a 132 nm-thick PFO film with 10.5 Ta$_2$O$_5$/SiO$_2$ (75 nm/51.20 nm) pairs on both sides. For metal mirrors, 1 nm Al (as seed layer) was deposited by electron beam physical vapor deposition and Ag was deposited by thermal evaporation in a vacuum chamber (Angstrom EvoVac) at a base pressure of $1 \times 10^{-7}$ mbar. Al was used as a wetting layer to improve percolation and optical quality of the thin Ag films. The Ag/DBR cavity contains a 95 nm-thick PFO film with 100 nm Ag on one side and 10.5 Ta$_2$O$_5$/SiO$_2$ pairs on the other. The Ag/Ag cavity contains a 170 nm-thick PFO film with 100 nm Ag on one side and 60 nm Ag on the other; this second-order cavity design for the Ag/Ag cavity was found to increase *Q*-factor in these cavities (see Supporting Information Figure S8e).

SiO$_2$ and Ta$_2$O$_5$ were deposited by radiofrequency magnetron sputtering at a base pressure of 10$^{-7}$ Torr, using 18 standard cubic centimeters per minute (sccm) Argon flow at 2 mTorr process pressure and 18 sccm Argon together with 4 sccm Oxygen flow at 4 mTorr process pressure for SiO$_2$ and Ta$_2$O$_5$, respectively. The additional oxygen flow during Ta$_2$O$_5$ deposition prevents the formation of unwanted sub-oxides. A layer of PFO was spin-coated using 24 mg mL$^{-1}$ PFO in toluene solution for the DBR/DBR cavity, 16 mg mL$^{-1}$ for the Ag/DBR cavity, 28 mg. mL$^{-1}$ for the Ag/Ag cavity, for 1 min at a speed of 2000 rpm, with an initial acceleration of 1000 rpm s$^{-1}$. Next, the sample was placed on a precision hotplate (Präzitherm, Gestigkeit GmbH) in an inert environment and the temperature was raised from 25 °C to 160 °C at a rate of approximately 30 °C min$^{-1}$. The upper temperature was then held for 10 min, followed by rapid quenching to room temperature by placing the sample on a metallic surface to induce the nematic phase Schlieren texture in the PFO film. Subsequently, approximately 15% $\beta$-phase fraction was induced in the by exposing the films to a saturated toluene vapour environment for 24 hours. The thicknesses of the films and thus of the active layers in the final cavity were controlled using a profilometer (Dektak, Bruker) on simultaneously prepared reference samples. Finally, the top DBR or Ag mirror was deposited following the same process described above.

*Angle-Resolved PL Measurements:* PL spectra were measured using Fourier imaging spectroscopy by imaging the back focal plane of a 40× objective (numerical aperture 0.75, Nikon Plan Fluor), set up on a commercial inverted microscope stand in reflection configuration (Nikon Eclipse Ti2-E). The third harmonic generation (THG) output of a diode-pumped Nd:YAG laser system (PL2210A, Ekspla), with wavelength 355 nm, repetition rate 250 Hz, and pulse duration 25 ps was used for excitation. Linear polarization of the pump was set by placing a Glan-Taylor polarizer (GT10-A, Thorlabs) before the objective. This polarization was rotated using a zero-order half wave plate (46-549, Edmund Optics). For the wide spot excitation, the diameter of the Gaussian beam at the sample plane was measured to be $d_{pump}$ ~ 150 µm, while for the small spot excitation, the diameter was reduced down to $d_{pump}$ ~ 3 µm to match the size of the localized emission centres in the Schlieren texture. The emitted light was directed towards the entrance of a spectrograph (Shamrock SR-500i-D2-SiL, Andor) equipped with an 1800 lines mm$^{-1}$ grating blazed at 500 nm and the PL spectra were imaged on an EM CCD camera (Newton 971, Andor) providing a spectral resolution of 40 pm (~ 300 µeV at 2.67 eV). The spatial coherence measurements presented in the Supporting Information were performed using a Michelson interferometer in the retro-reflector configuration and the resulting interferograms were imaged on an sCMOS camera (ORCA-Flash 4.0, Hamamatsu).

**Acknowledgements**

The authors thank Prof. Donal Bradley and the Sumitomo Chemical Company for provision of PFO. F.L.R acknowledges funding from the Alexander von Humboldt Foundation through a Humboldt Fellowship. A.M. acknowledges funding from the European Union Horizon 2020 research and innovation programme under the Marie Skłodowska-Curie grant agreement No. 101023743 (PolDev). This research was financially supported by the Alexander von Humboldt Foundation (Humboldt Professorship to M.C.G.).

**Supporting Information**

**Schlieren texture induced Anderson localization in an organic exciton-polariton laser**

*Florian Le Roux*[1*], *Andreas Mischok*[1]*, Francisco Tenopala-Carmona*[1], *Malte C. Gather*[1,2*]

[1] Humboldt Centre for Nano- and Biophotonics, Department of Chemistry, University of Cologne, Greinstr. 4-6, 50939 Köln, Germany

[2] Organic Semiconductor Centre, SUPA School of Physics and Astronomy, University of St Andrews, St Andrews, KY16 9SS, UK

# I - Optical constants for aligned PFO β-phase

The optical constants and PL spectrum of macroscopically aligned 15% β-phase PFO depicted in **Figure S1**a, b show a strong preferential alignment of the polymer along the y-direction and a well-resolved vibronic progression of the emission from β-phase PFO, respectively emission.

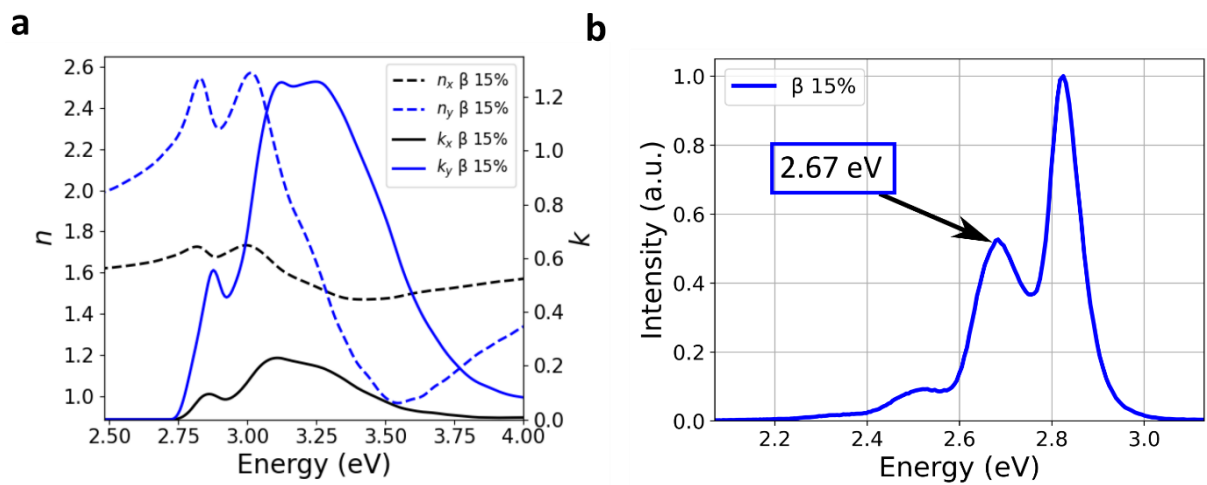

**Figure S1. Photophysical properties of aligned PFO β-phase. a**, In-plane optical constants of macroscopically aligned 15% β-phase PFO along the direction of alignment (y, blue lines) and perpendicular to the alignment (x, black lines), showing the extinction coefficient, $k_x$ and $k_y$ (solid lines), and refractive index, $n_x$ and $n_y$ (dashed lines). **b**, PL spectrum for a spin-coated and aligned PFO thin film containing 15% *β*-phase. (Reproduced with permission from Ref. 3.)

## II - Strong localization of polaritons under polarized excitation

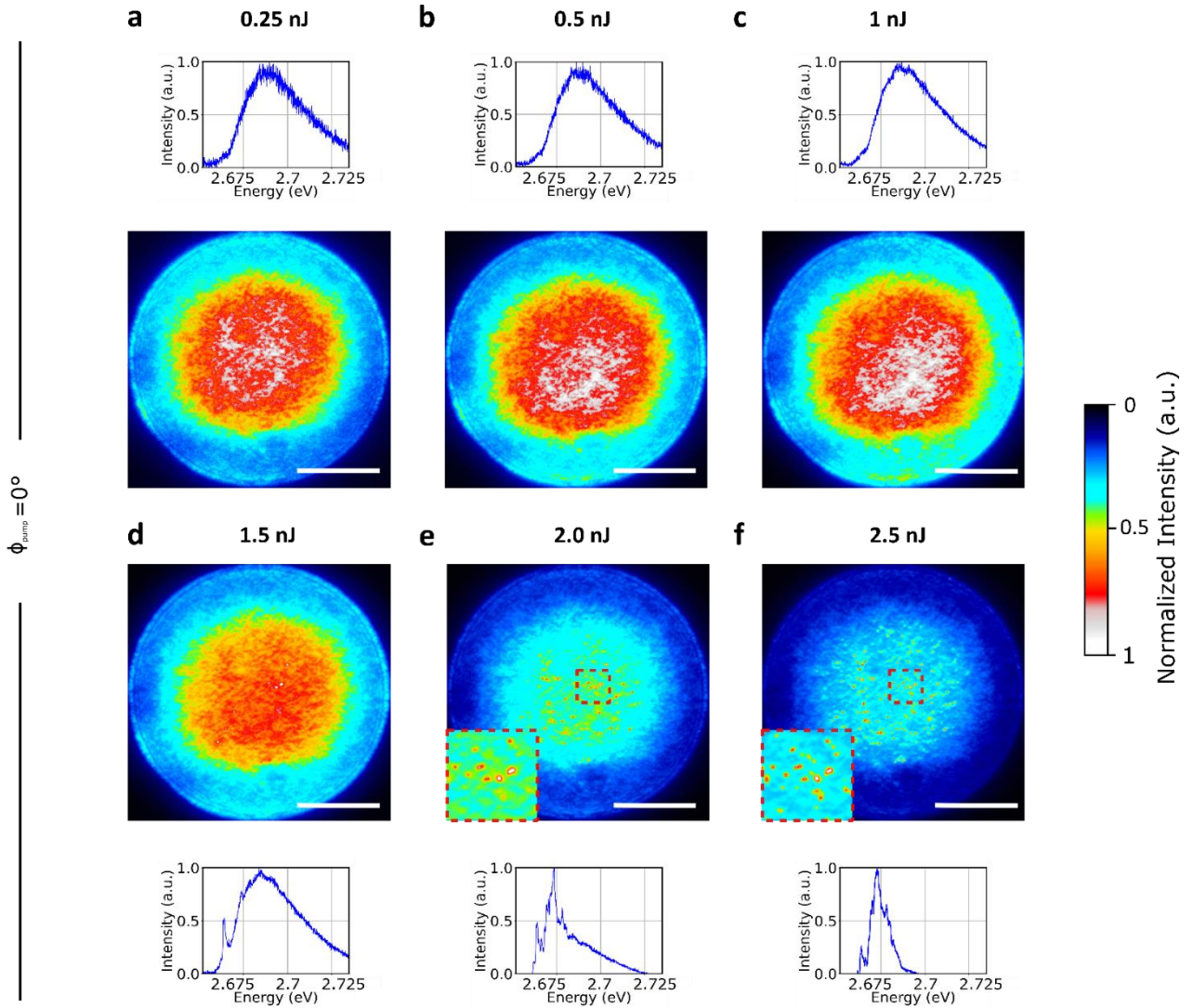

**Figure S2. Spatially and spectrally resolved emission from the Schlieren textured DBR/DBR cavity under non-resonant optical pumping with $\phi_{pump} = 0°$.** Spatially and spectrally resolved emission from the DBR/DBR cavity for increasing excitation pulse energies; a) 250 pJ, b) 500 pJ, c) 1 nJ, d) 1.5 nJ, e) 2 nJ, f) 2.5 nJ per pulse. The insets in e and f show magnifications of the red dashed rectangles in the corresponding main panels. Scale bars: 50 μm.

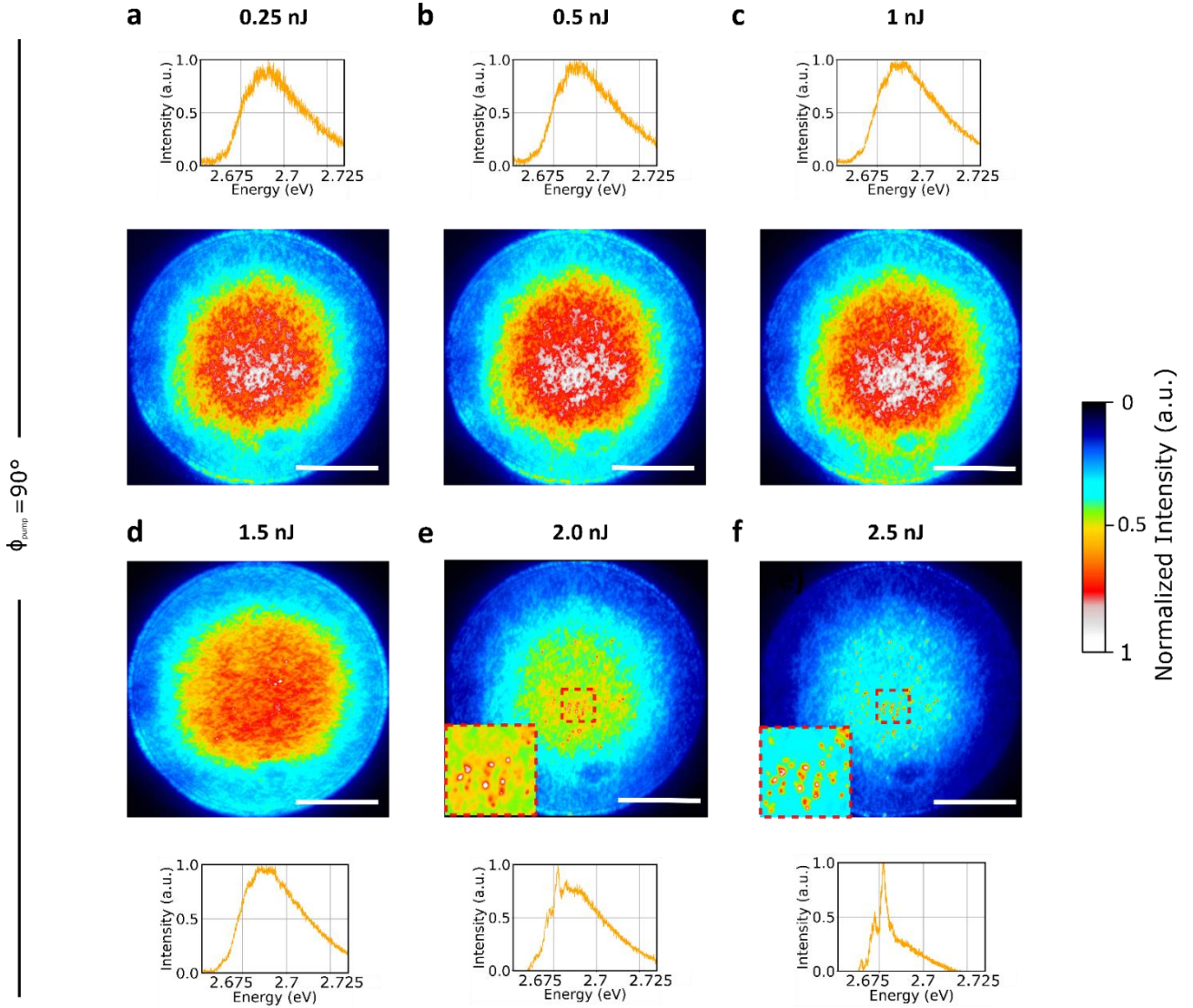

**Figure S3. Spatially and spectrally resolved emission from the Schlieren textured DBR/DBR cavity under non-resonant optical pumping with $\phi_{pump}= 90°$.** Spatially and spectrally resolved emission from the DBR/DBR cavity for increasing excitation pulse energies; a) 250 pJ, b) 500 pJ, c) 1 nJ, d) 1.5 nJ, e) 2 nJ, f) 2.5 nJ per pulse. The insets in e and f show magnifications of the red dashed rectangles in the corresponding main panels. Scale bars: 50 μm.

# III - Strong localization of polaritons in the Schlieren texture

**Local Transition Dipole Moment Orientation**

Finite-difference time-domain (FDTD) simulations of the in-plane component of $|\boldsymbol{E}|^2$ were performed to reproduce the strong localization of exciton-polaritons inside the active layer. The simulations were performed using the FDTD 3D Electromagnetic Simulator from Lumerical-Ansys[1].

**Figure S4**a shows the vertical structure of the DBR/DBR cavity used for the simulation. The in-plane simulated surface was 13.97 μm wide by 13.97 μm long, in correspondence with the optical polarized micrograph shown in Figure S3b. The spatially-resolved PL obtained using a 150 μm wide polarized excitation spot with a pulse energy of 1.5 nJ and $\phi_{\text{pump}}= 0°$ is also shown in Figure S4c with the high intensity emission centre clearly visible.

It has been shown that the combination of cross-polarized microscopy and polarized PL is an efficient method to lift the inherent degeneracy that exist between orthogonal directions in cross-polarized microscopy[2]. PL measurements represented in Figure S4d for $\phi_{\text{pump}}= 0°$ (left, red) and $\phi_{\text{pump}}= 90°$ (right, blue) at an excitation energy of 500 pJ were therefore recorded in order to delineate the orientation of the local transition dipole moments.

Figure S4e shows a comparison between the emission intensities obtained for $\phi_{\text{pump}} = 0°$ (red) and $\phi_{\text{pump}} = 90°$ (blue). Since both images are 8-bit grayscale, the emission intensity of each pixel $I_{x,y}$ is an integer between 0 and 255. The pixel color $C_{x,y}$ in Figure S3e is obtained as follows. If $I_{x,y\text{-}0°} > 130$ and $I_{x,y\text{-}90°} < 130$, $C_{x,y}$ is set to red, which means that this location ($x,y$) is dominated by emission from transition dipole moments closer to 0 than 90° (we call this red domain the 0° domain in the following). Conversely, if $I_{x,y\text{-}0°} < 130$ and $I_{x,y\text{-}90°} > 130$, $C_{x,y}$ is set to blue, which means that this location ($x,y$) is dominated by emission from transition dipole moments closer to 90° than 0°(we call this blue domain the 90° domain in the following). Finally if both $I_{x,y\text{-}0°} > 130$ and $I_{x,y\text{-}90°} > 130$, $C_{x,y}$ is green and the pixel is not clearly dominated by either orientation.

Domains resulting from this procedure are mostly complimentary with large regions dominated by either the 0° or 90° orientation. The final delimitation between 0° or 90° domains is obtained by comparing the domains in Figure S3e to the polarized micrograph in Figure S3b which enables the attribution of the green domains to either the 0° or 90° domains. The left-most panel in Figure S4f shows the final 0° and 90° domains.

The local transition dipole moment orientation $T_{x,y}$ is then calculated using the intensity of the corresponding pixel $M_{x,y}$ in the polarized micrograph: a dark state with 0 intensity is observed for orientation of the local transition dipole moment parallel to either the polarizer or analyzer, a bright state with 255 intensity is observed when the transition dipole moment is at 45° relative to either the polarizer or the analyzer. A simplified linear procedure for calculating $T_{x,y}$ is then if (x,y) belongs to the 0° domain, then $T_{x,y}$ belongs to [0°,45°] and $T_{x,y} = \frac{45 \times M_{x,y}}{255}$. Conversely if (x,y) belongs to the 90° domain, then $T_{x,y}$ belongs to [45°,90°] and $T_{x,y} = 90 - \frac{45 \times M_{x,y}}{255}$. The final orientations of $T$ are represented by black arrows and shown in the two panels of Figure S1g.

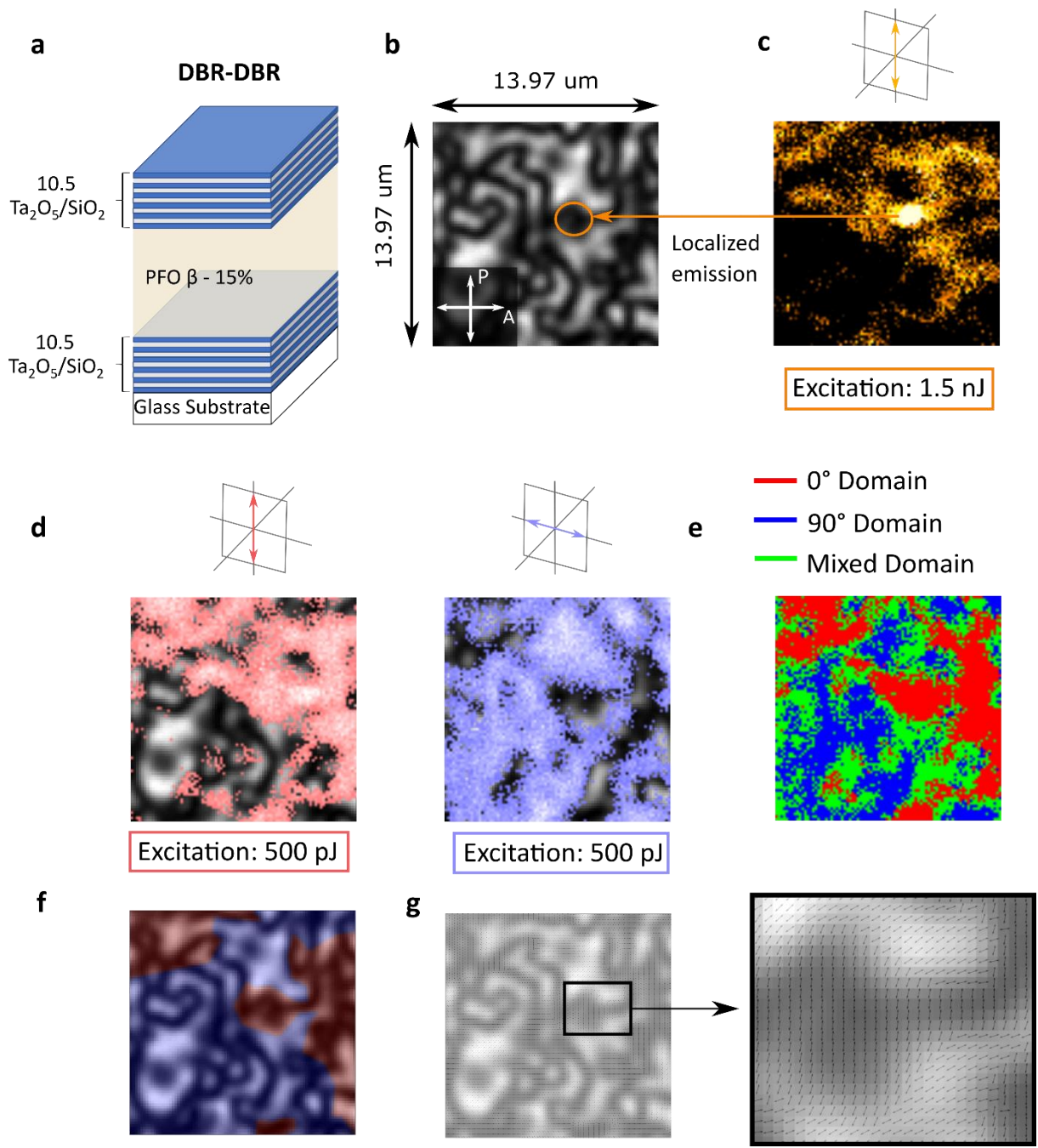

**Figure S4. Setting up of the FDTD simulation. a**, Schematic of the DBR/DBR cavity used as model in the FDTD simulation. **b**, Experimental polarized optical micrograph of the surface of the DBR/DBR cavity recorded with the sample placed between the crossed polarizer (P) and analyzer (A) pair. No light is transmitted when the exciton transition dipole moment lies either parallel to the polarizer or the analyzer, while maximum transmission occurs when the transition dipole moment lies at 45° relative to both the analyzer and polarizer. **c**, Real-space PL image obtained using a 150 μm wide polarized excitation spot with a pulse energy of 1.5 nJ and $\phi_{pump}= 0°$. The high intensity spot is clearly visible and its position in b is marked by an orange circle. **d**, Real-space emission images obtained using $\phi_{\text{pump}} = 0°$ (red, left) and $\phi_{\text{pump}} = 90°$ (blue, right) at 500 pJ. The images are overlayed on the polarized optical micrograph from b. **e**, Comparative image obtained following the procedure described in the text showing 0° (red), 90° (blue) or mixed (green) domains. **f**, 0° (red), 90° (blue) domains following comparison of e with the domains of the polarized micrograph in b. **g**, Black arrows representing the local transition dipole moment orientation. The panel on the right shows a zoomed-in version of this image. Both images are overlayed on the polarized optical micrograph from b.

**FDTD simulations**

The FDTD simulation is passive in the sense that no re-emission is considered once the electric field has been absorbed by one of the elements of the simulation. The optical constants used for the different materials ($Ta_2O_5$, $SiO_2$, aligned 15% β-phase PFO) were reproduced from Ref. 3. The active layer was defined as a liquid-crystalline layer in which the local transition dipole moment $T_{x,y}$ can be set by applying the corresponding transformation to the refractive index of the layer.

The simulation monitors the in-plane component of $|E|^2$ inside the active layer for the first 2500 fs. The so-called active region where initial emission is permitted is a 7 μm x 7 μm square in the center of the 13.97 μm x 13.97 μm region of interest to avoid border aberrations that can arise with perfectly matched layers (PMLs) boundary conditions. The emission region is represented by an orange rectangle in Figure S1a. Each pixel was given an initial emission power $P_{x,y}(t=0)$ proportional to the square of the dot product between the transition dipole moment orientation $T_{x,y}$ and the excitation polarization ($\phi_{\text{pump}} = 0°$), simulating initial absorption inside the cavity. Each dipole is given a random phase to prevent interference and broad spectral characteristics resembling a thin-film of PFO in its β-phase, i.e., center energy at 2.67 eV to match the (0-1) vibronic peak of the emission, pulse length of 10 fs and bandwidth of 183 meV.

The resulting $|E|^2$ inside the active layer at multiple time points between 0 fs and 2496 fs is shown in **Figure S5**. Strong localization of the emission is evident in Figure S4b and matches the high-intensity PL spot of Figure S4c. The combination of cross-polarized microscopy, polarized PL and FDTD simulation forms a precise tool for the characterization of the strong localization of the exciton-polaritons inside the structure. The simulation also confirms that the mode surface contributing to the mode volume $V_{\text{sim}} = h_{\text{eff}} \times S_{\text{sim}}$, where $h_{\text{eff}}$ is an effective height which mainly depends on the penetration depth through the mirrors, is reduced compared with aligned active layers[3]: $S_{\text{Aligned}}$ ~20 μm$^2$ compared with $S_{\text{Sim}}$ ~7 μm$^2$, a nearly three-fold reduction in mode volume between the aligned and textured cavities.

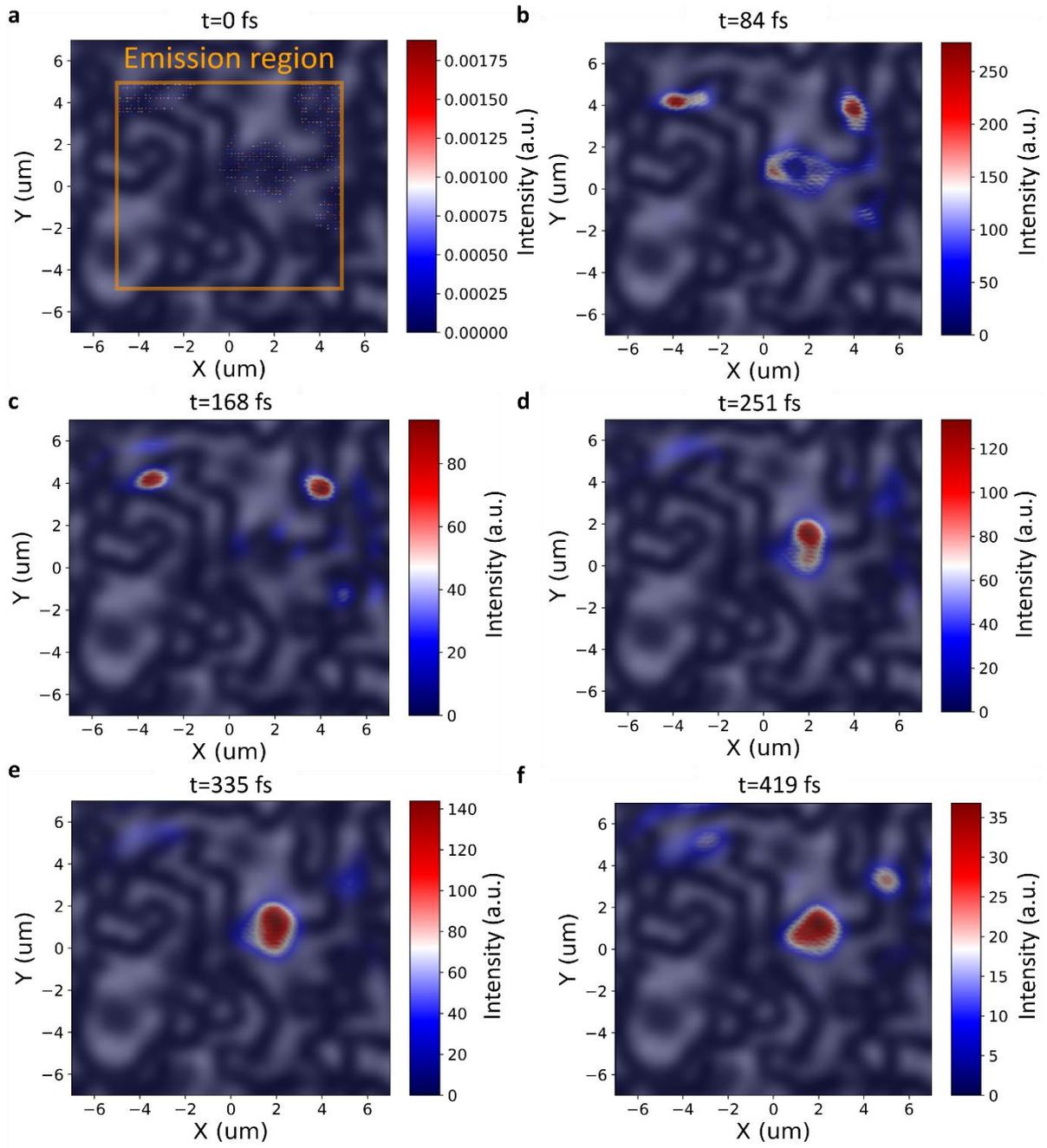

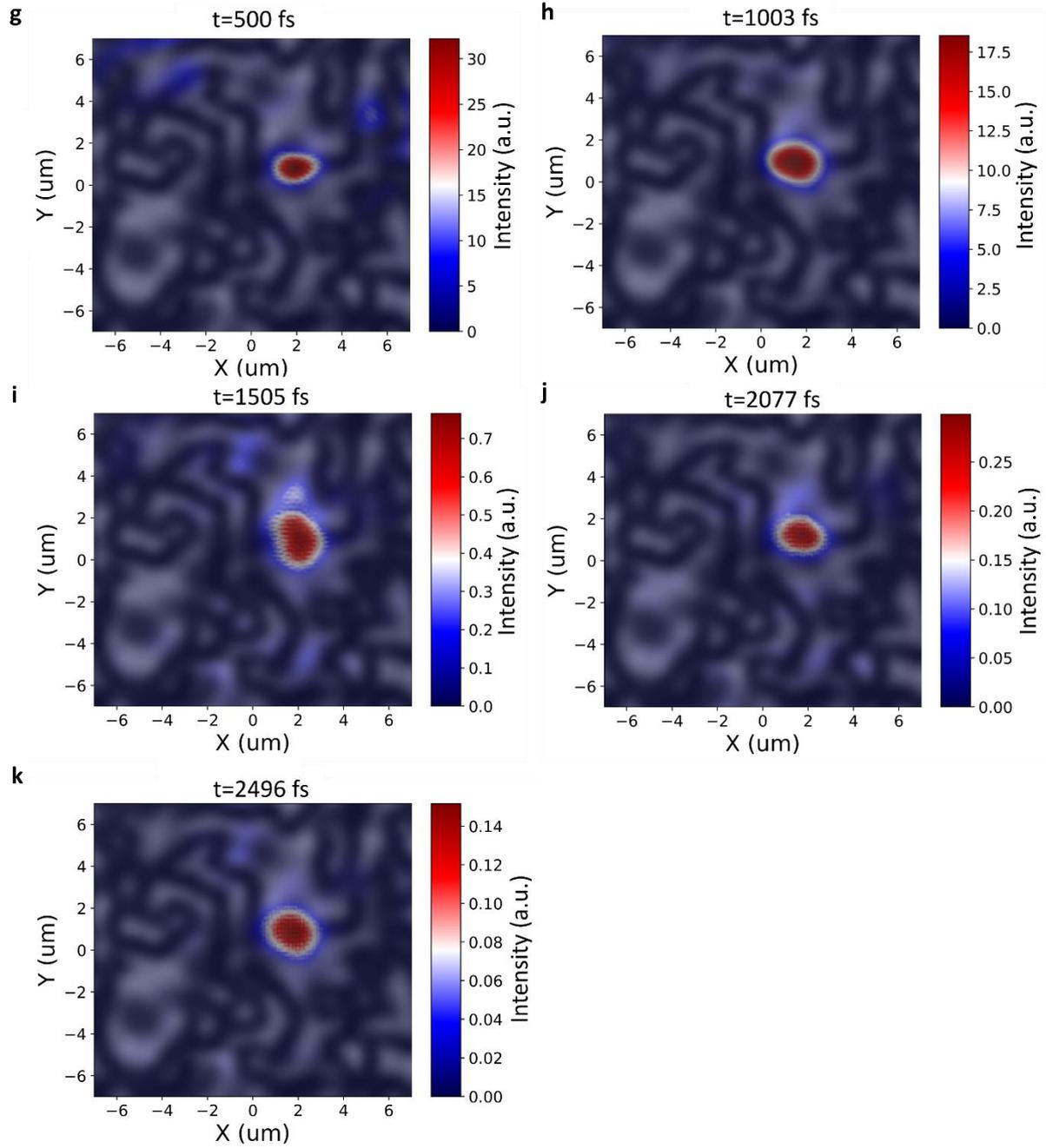

**Figure S5. In-plane intensity simulated FDTD.** Two dimensional FDTD calculations of in-plane $|E|^2$ at times a) t = 0 fs, b) t = 84 fs, c) t = 168 fs, d) t = 251 fs, e) t = 335 fs, f) t = 419 fs, g) t = 500 fs, h) t = 1003 fs, i) t = 1505 fs, j) t = 2077 fs and k) t = 2496 fs. The electric field intensity is plotted on top of the polarized optical micrograph shown in Figure 1a. The simulations were performed following the procedure described in the text. The region inside the orange rectangle corresponds to the initial emission region. Strong localization of the electric field intensity is evident and confirms the experimental observation in Figure 1b.

## IV - Kinetic Model for the Aligned and Non-Aligned Cavities

The input-output lasing curves were analysed using a kinetic model derived from the Gross-Pitaevskii equation[4]. The resulting master equation for the polariton population, $n_{LP}(t)$, and exciton reservoir population, $n_R(t)$, is:

$$\frac{dn_R}{dt} = \left(1 - \frac{n_R}{N_0}\right) P(t) - \frac{n_R}{\tau_R} - k_B n_R^2 - \frac{W_{ep}}{d} n_R n_{LP} \qquad (1)$$

$$\frac{dn_{LP}}{dt} = W_{ep} n_R n_{LP} - \frac{n_{LP}}{\tau_{LP}} + df \frac{n_R}{\tau_R} \qquad (2)$$

where $N_0$ is the bare excitation density inside a PFO film ($\sim 7 \times 10^{20}$ cm$^{-3}$), $\tau_R$ the reservoir exciton lifetime ($\sim 450$ ps), $d$ the active layer thickness, $\tau_{LP}$ the LP lifetime. The pump term $P(t)$ is a Gaussian pump term, i.e. $P(t) = o_a P_{eff} \exp(-t^2(2\sigma^2)^{-1})$ with $FWHM = 2\sqrt{2 ln 2}\sigma = 25$ ps and $P_{eff} = P_0 (dS\hbar\omega_{pump})^{-1}$ where $S$ is the pumped surface and $\hbar\omega_{pump} = 3.49$ eV. The free parameters in the model are the exciton bimolecular annihilation rate, $k_B$, the exciton reservoir-to-LP resonant scattering rate, $W_{ep}$, and the fraction of spontaneous scattering from the exciton reservoir to the LP $f$. Additionally, the term $o_a$ corresponds to the pumping of the exciton reservoir population according to the dot product between pump polarization and transition dipole moment $o_a = \frac{2\|\boldsymbol{\mu}.\boldsymbol{E}\|^2}{\|\boldsymbol{\mu}\|^2 \|\boldsymbol{E}\|^2}$.

The experimental input-output curves for a non-aligned 15% β-phase PFO cavity and an aligned 15 % β-phase PFO cavity and the corresponding fits to the kinetic model are shown in **Figure S6** (reproduced with permission from Ref. 3), and results for the aligned cavity are displayed in **Table S1**.

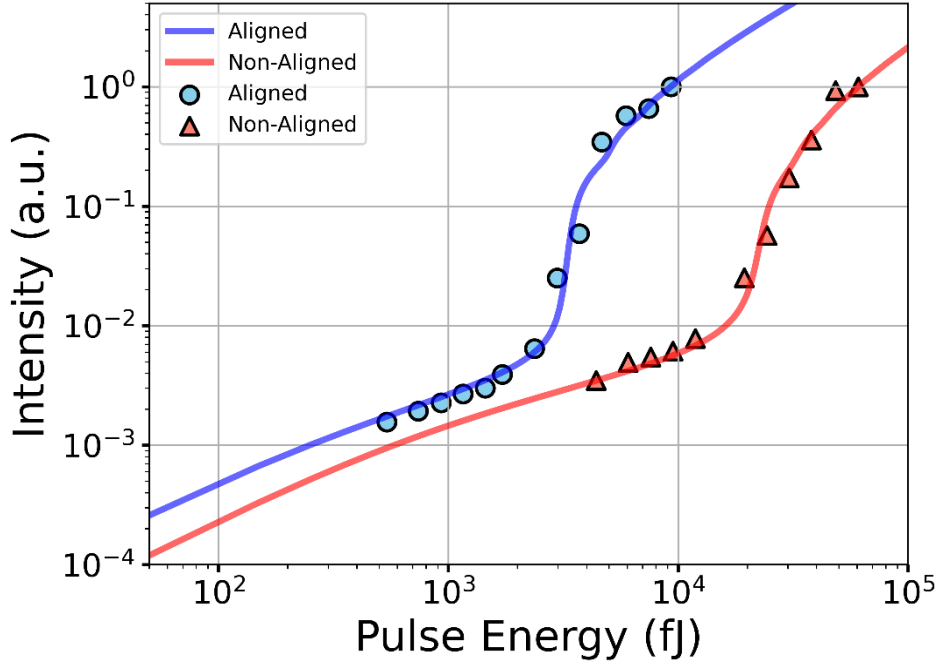

**Figure S6. Emission intensity for the aligned and non-aligned non-textured cavities.** Integrated emission intensity versus incident excitation pulse energy for a DBR/DBR cavity containing a non-aligned (red symbols) and an aligned (blue symbols) layer of 15% β-phase PFO. The data was reproduced with permission from Ref. 3.

| Sample | Pre-factor $o_a$ | Active layer thickness [nm] | Bimolecular annihilation rate $k_B$ [cm$^3$s$^{-1}$] | Resonant polariton scattering rate $W_{ep}$ [cm$^3$s$^{-1}$] | Spontaneous scattering fraction $f$ | LP radiative decay time from peak width $\tau_{LP-exp}$ [fs] | Incident threshold pulse energy $P_{th}$ |
|---|---|---|---|---|---|---|---|
| Non-Aligned (DBR/DBR) | 1 | 155 | 7.5 x 10$^{-9}$ | 1.2 x 10$^{-6}$ | 0.080 | 132 | 14.50 pJ |
| Aligned (DBR/DBR) | 2 | 130 | 7.5 x 10$^{-9}$ | 2.6 x 10$^{-6}$ | 0.031 | 132 | 2.23 pJ |

**Table S1.** Polariton lasing performance and parameters extracted from the kinetic model for the non-aligned and aligned 15% β-phase PFO cavities (reproduced with permission from Ref. 3).

**V- Spatial Coherence of the Emission**

Spatial coherence of the emission above threshold is a signature of polariton lasing. **Figure S7** shows the interferometry measurements performed on the different cavities using a Michelson interferometer in the retroreflector configuration[6,7]. The images to the left of the "+" sign show an example of the recorded image obtained by blocking one of the two arms of the interferometer, the images to the right of the "+" sign and thus to the left of the "=" sign show the mirror image obtained by freeing the blocked arm and blocking the other arm of the interferometer. Finally, the images to the right of the "=" sign displays clear fringes obtained through interference of the two paths by having both arms freed, showing clear spatial coherence above threshold.

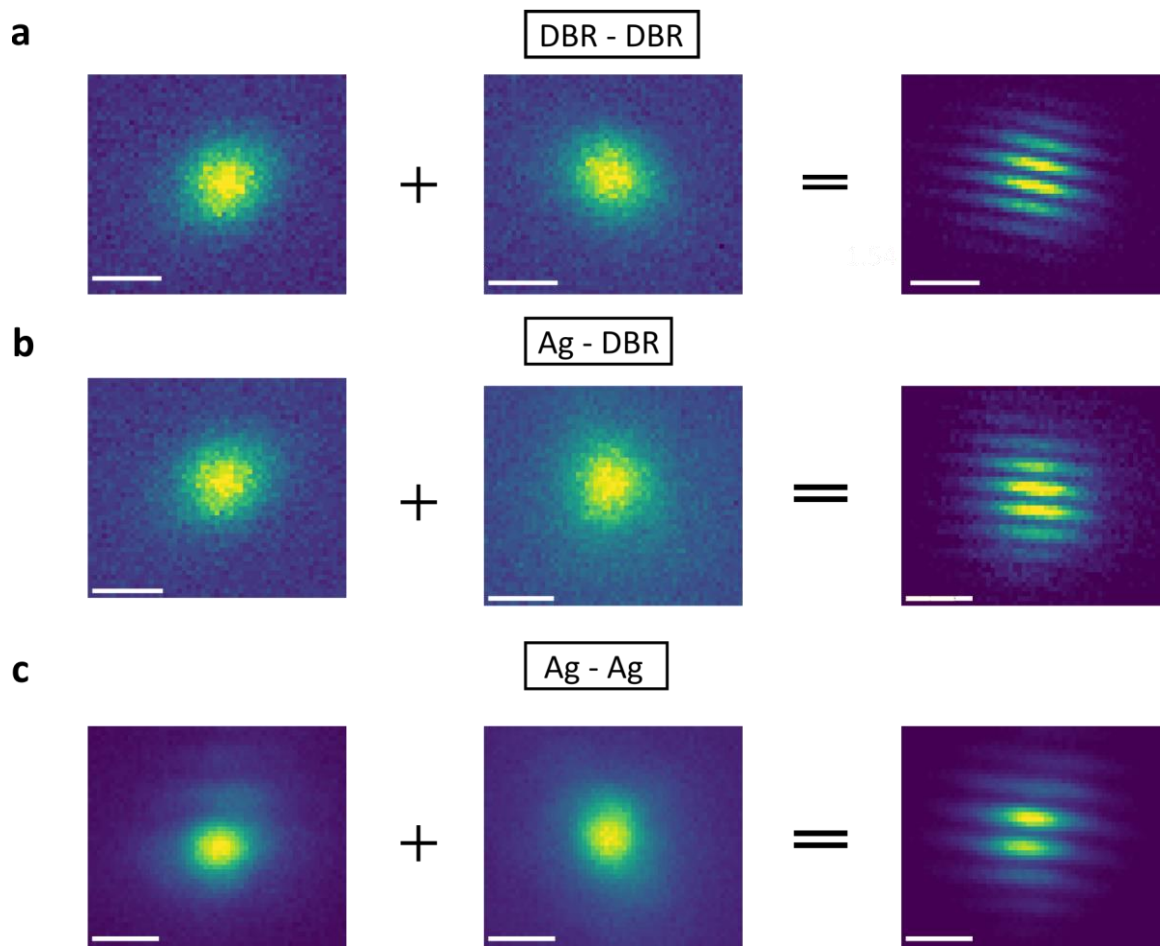

**Figure S7. Spatial coherence of lasing emission from different cavities.** Interferometry measurements performed on a) the DBR/DBR, b) the Ag/DBR, c) and the Ag/Ag cavities using a Michelson interferometer in the retroreflector configuration[6,7]. Scale bars: 2,5 μm.

# VI - Transfer matrix calculations and the effect of roughness on the Q-factor

When nominal mirror reflectivity is continuously increased, the *Q*-factor achievable in optical microcavities made from such mirrors usually saturates at some point due to losses from fabrication limitations, specifically film roughness, which can typically reach a few nm[44] for spun polymer films. TMCs in **Figure S8**a demonstrate that as little as 1 nm roughness can account for a 4-fold reduction in the *Q*-factor of an aligned DBR/DBR cavity. Interestingly, the importance of roughness decreases as the top Ag mirror is added (with a 1 nm roughness accounting for a more modest 21% *Q*-factor reduction for the Ag/DBR relative to the ideal case), as the linewidth is already broadened by the presence of the lossy metal mirror (Figure S8d), suggesting the Ag/DBR is less prone to fabrication imperfections than the DBR/DBR cavity. The increase in *Q*-factor for the DBR/DBR and Ag/DBR cavities from the ideal values of $Q_{DBR/DBR}$ = 2661 (Figure S8a) and $Q_{Ag/DBR}$ = 200 (Figure S8e) up to the experimental $Q_{DBR/DBR-exp}$ = 5300 (Figure S8c) and $Q_{Ag/DBR-exp}$ = 333 (Figure S8d) can be interpreted as the *Q*-factor enhancement induced by strong localization inside the Schlieren texture and agrees with previous reports. Losses increase when the top Ag mirror is evaporated directly on the polymer layer as the metal film formation becomes prone to cracks and imperfections leading to $Q_{Ag-Ag-exp}$ = 67 (Figure S8i); these losses are found to be equivalent to a simulated increase in roughness of the polymer film of the order of 5 nm which gives $Q_{Ag-Ag}$ = 77 (Figure S8h). $Q_{Ag-Ag-exp}$ is however still higher relative to the *Q*-factor of an ideal Ag/Ag cavity with a first-order design (Figure S8j). For these cavities, the ideal *Q*-factor value decreases to 53 and no lasing could be observed from such structures.

The corresponding transfer matrix calculations (TMC)s are detailed below. First the results for a DBR/DBR cavity containing a 132 nm-thick aligned layer of PFO with no roughness was simulated. The results are shown in Figure S8a. The broadening of the LP is expected to be mainly due to homogeneous broadening[9] with a corresponding Lorentzian of width 1 meV corresponding to a Q-factor of 2661. Experimental observations[3] show that this value is not reached in an actual cavity, instead, the experimental *Q*-factor is closer to 500 – 600 and the measured linewidth of the LP is approximately 5 meV.

We suggest that the main source of losses is the roughness of the polymer film as AFM measurements[8] indicate a roughness of at least 1 nm for typical LCCP films. TMCs simulating a variation in roughness were performed taking the average of 10 separate simulations where for *i* in [1,10] $d(i) = 132 + i \times 0.1$ nm. As *i* increases, the position of the LP red-shifts with the resulting average reflectivity $R = \sum \frac{R_i}{10}$ shown in Figure S8b. The LP mode was fitted using a Gaussian with linewidth 4.02 meV ($Q = 662$) demonstrating that roughness can easily account for a 4-fold increase in linewidth and corresponding reduction in $Q$-factor.

Similar calculations were performed for the Ag/DBR cavity. Figure S8d shows the TE-reflectivity spectrum obtained without any roughness consideration, yielding a fitted linewidth of 13.33 meV ($Q \sim 200$). Figure S8d shows the result for a 1 nm roughness with a linewidth increase to 16.78 meV and a $Q$-factor reduction down to 158. Compared with the DBR/DBR cavity, the roughness this time only accounts for a 1.25-fold increase in linewidth as the Ag optical losses now represents the main source of losses.

TMCs were also performed for the Ag/Ag cavity. Figure S8g shows the TE-reflectivity spectrum without any roughness consideration; the fitted linewidth is 21.2 meV and $Q = 125$. Including 5nm roughness, the $Q$-factor decreases to 77.

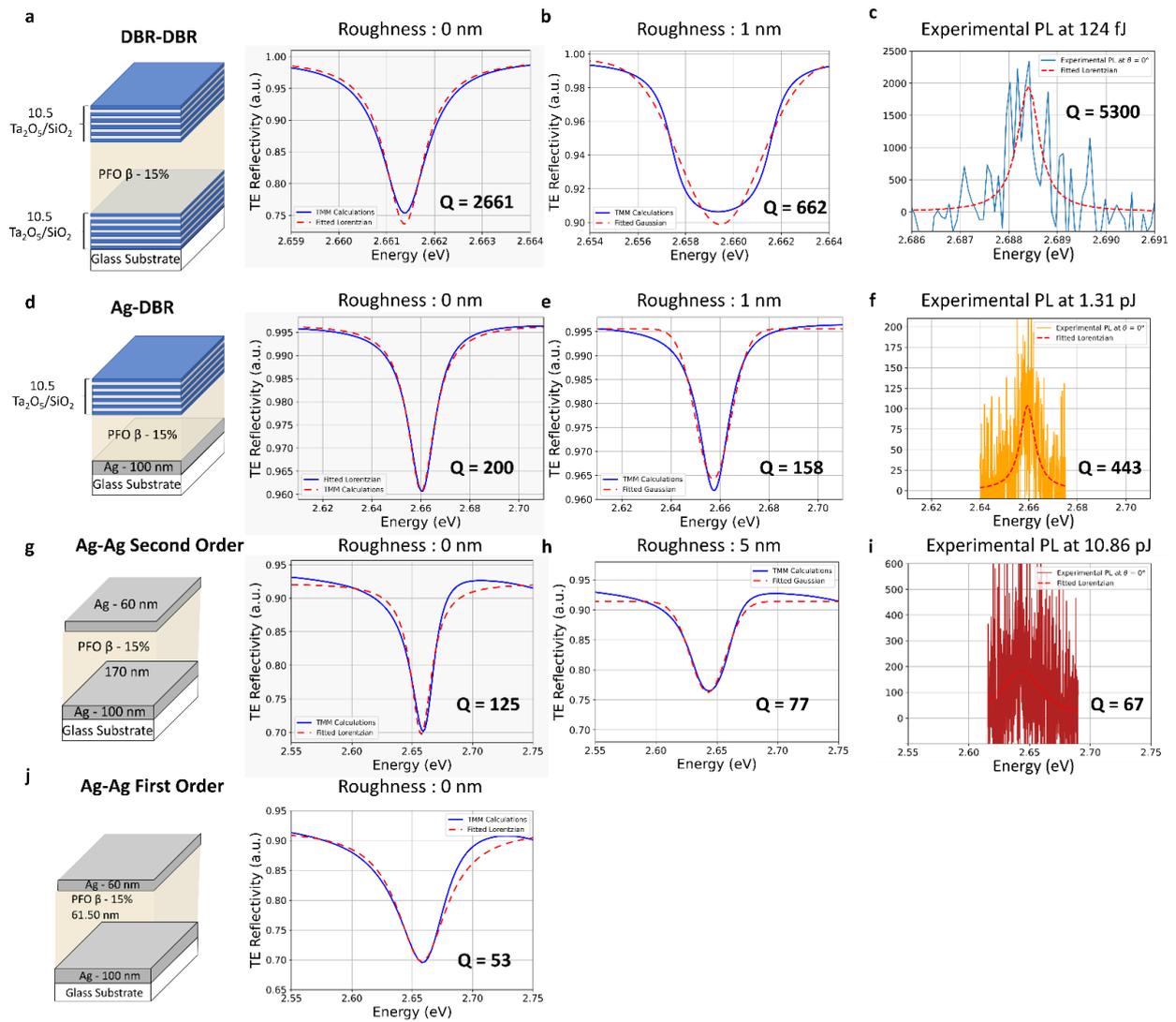

**Figure S8. Q-Factor comparison.** Schematic illustrations of the DBR/DBR (a), Ag/DBR (d) and Ag/Ag second order (g), and Ag/Ag first order (j) cavities used as models in the transfer matrix calculations. Directly to the right of each schematic are the TE-reflectivity spectra obtained from TMC assuming no roughness. b), e) and h) show the corresponding TE-reflectivity spectra after including a 1 nm (c,f) and 5 nm (h) roughness, again obtained via TMC. c), f), i) show the experimentally recorded emission spectra below threshold for θ = 0°.